\documentstyle[12pt]{article}

\catcode`\@=11

\newcount\hour
\newcount\minute
\newtoks\amorpm
\hour=\time\divide\hour by 60
\minute=\time{\multiply\hour by 60 \global\advance\minute by-\hour}
\edef\standardtime{{\ifnum\hour<12 \global\amorpm={am}%
        \else\global\amorpm={pm}\advance\hour by-12 \fi
        \ifnum\hour=0 \hour=12 \fi
        \number\hour:\ifnum\minute<10
        0\fi\number\minute\the\amorpm}}
\edef\militarytime{\number\hour:\ifnum\minute<10
0\fi\number\minute}

\def\draftlabel#1{{\@bsphack\if@filesw {\let\thepage\relax
   \xdef\@gtempa{\write\@auxout{\string
      \newlabel{#1}{{\@currentlabel}{\thepage}}}}}\@gtempa
   \if@nobreak \ifvmode\nobreak\fi\fi\fi\@esphack}
        \gdef\@eqnlabel{#1}}
\def\@eqnlabel{}
\def\@vacuum{}
\def\marginnote#1{}
\def\draftmarginnote#1{\marginpar{\raggedright\scriptsize\tt#1}}
 \def \lc {light-cone\ }

\def\draft{
        \pagestyle{plain}
        \overfullrule=2pt
        \oddsidemargin -.5truein
        \def\@oddhead{\sl \phantom{\today\quad\militarytime} \hfil
        \smash{\Large\sl DRAFT} \hfil \today\quad\militarytime}
        \let\@evenhead\@oddhead
        \let\label=\draftlabel
        \let\marginnote=\draftmarginnote
        \def\ps@empty{\let\@mkboth\@gobbletwo
        \def\@oddfoot{\hfil \smash{\Large\sl DRAFT} \hfil}
        \let\@evenfoot\@oddhead}
        \def\@eqnnum{(\theequation)\rlap{\kern\marginparsep\tt\@eqnlabel}%
        \global\let\@eqnlabel\@vacuum}  }

\newcommand{\rf}[1]{(\ref{#1})}
\renewcommand{\theequation}{\thesection.\arabic{equation}}
\renewcommand{\thefootnote}{\fnsymbol{footnote}}

\newcommand{\newsection}{    
\setcounter{equation}{0}
\section}
\def\appendix#1{
  \addtocounter{section}{1}
  \setcounter{equation}{0}
  \renewcommand{\thesection}{\Alph{section}}
  \section*{Appendix \thesection\protect\indent \parbox[t]{11.15cm} {#1} }
  \addcontentsline{toc}{section}{Appendix \thesection\ \ \ #1}
  }

\def\x'{\hbox{\'x}}
\def\y'{\hbox{\'y}}

\def\X'{\hbox{\'X}}

\def\al{{\lambda}}

 \def \const {{\rm const}}

\def\vm{{\mu}}
\def\vn{{\nu}}

\def\sfa{{\sf a}}

\def\dsfa{\dot{\sf a}}
\def\dsfb{\dot{\sf b}}

\def \td {\tilde }

\def \foot {\footnote}
\def \bi{\bibitem}
\def \la {\label}

\def \ha {{1 \over 2}}
\def \ep{\epsilon}

\def \ov {\over}

\def\nline{\,\nabla\kern -0.7em\raise0.2ex\hbox{/}\,\,}
\def\yline{\,y\kern -0.47em /}
\def\aline{\,a\kern -0.49em /}
\def\parline{\,\partial\kern -0.55em /\,\,}

\def \t {\theta}
\def \s{\sigma}

\def \del{\partial}
\def \m {M}
\def \n {N}

\def\det{\hbox{det}}
\def\be{\begin{equation}}
\def\ee{\end{equation}}

\def \ci {\cite}
\def \g {\gamma}

\def \k {\kappa}
\def \l {\lambda}

\def \del{\partial}

\def\det{\hbox{det}}
\def\be{\begin{equation}}
\def\ee{\end{equation}}

\def \ci {\cite}
\def \g {\gamma}

\def \k {\kappa}
\def \l {\lambda}

\def \m {\mu}
\def \n {\nu}

\def\x'{\mathaccent 19 x}
\def\y'{\mathaccent 19 y}
\def\n'{\mathaccent 19 n}
\def\u'{\mathaccent 19 u}

\def\X'{\mathaccent 19 X}
\def\Y'{\mathaccent 19 Y}
\def\Z'{\mathaccent 19 Z}

\def\et'{\mathaccent 19 \eta}
\def\th'{\mathaccent 19 \theta}
\def\lam'{\mathaccent 19 \lambda}
\def\varet'{\mathaccent 19 \vartheta}
\def\rh'{\mathaccent 19 \rho}
\def\ph'{\mathaccent 19 \phi}
\def\xb'{\mathaccent 19 {\bar{x}}}
\def \adss {$AdS_5 \times S^5$\ }
\def \N {{\cal N}}
\def \lc {light-cone\ }
\def \ta { \tau}
\def \s { \sigma }

\def \sg {\sqrt {g }}
\def \te {\theta}
\def \vp {\varphi}
\def \gij {g_{ab}}
\def \xp {x^+}
\def \xm {x^-}
\def \p {\phi}
\def \vt {\theta}
\def \bx {\bar x} \def \a { \alpha}
\def \r {\rho}
\def \fourth {{1 \ov 4}}
\def \DD {{\cal D}}
\def \half {{1 \ov 2}}
\def \inv {^{-1}}
\def \ri {{\rm i}}
\def \D {{\cal D}}
\def \DD {{\rm D}}
\def \vr {\varrho}
\def \diag {{\rm diag}} \def \td { \tilde }
\def \tta {\td \eta}
\def \cA {{\cal A}}
\def \cB   {{\cal B}}
\def \na {\nabla}
\def \al {\perp}
\def \PP {{\cal P}} 

\def \po {\p_0}
\def \TT {{\cal T}} 
\def \T {{\rm  T}}

\def \z {\zeta}

\begin{document}

\begin{titlepage}
\vspace{2 cm}

\begin{center}
{\LARGE
On limits of   superstring in  AdS$_5 \times $S$^5$ }\\[.2cm]
\vspace{1.1cm}
{\large 
A.A. Tseytlin\footnote{\ Also at 
 Lebedev Physics Institute, Moscow} }\\

\vspace{18pt}
 {\it
 Blackett Laboratory, Imperial College   \\
London, SW7 2BZ, U.K. \\
}
\vspace{6pt}

\vspace{18pt}
 {\it
 Department of Physics,
The Ohio State University  \\
Columbus, OH 43210-1106, USA\\
}

\end{center}

\vspace{2cm}

\begin{abstract}
The superstring action in  \adss  depends on two parameters:
the inverse string tension $\a'$ and the radius $R$. 
The ``standard''  AdS/CFT correspondence requires that the string coordinates 
are rescaled so that the action depends only on one combination of the two: 
$\sqrt \lambda = R^2/\a'$. Then  $\lambda \to 0$ limit is equivalent
to $R \to 0$ for fixed $\a'$ or to the zero-tension limit in \adss:
$ \a' \to \infty$ for fixed $R$.
After reviewing  previous work on the
light cone superstring  we explicitly obtain the $\lambda= 0$ 
form of its action. 
Its zero-mode part is the same as  the superparticle action in
$AdS_5 \times S^5$,  and thus the $\lambda=0$ string spectrum  must include, 
as expected, the  ``protected'' type IIB supergravity states. 
Following recent suggestions, it is conjectured  that the  spectrum of this
tensionless string  should as well  contain higher spin massless states in $AdS_5$. 
We  also discuss the case of another parametrization 
of the string action which has straightforward $R\to \infty$ flat space limit
but  where $R \to 0 $  and $\a' \to \infty $ limits are not equivalent. 
There  $R \to 0$ corresponds  to shrinking $S^5$  to zero the size  
and  ``freezing'' the fluctuations of the radial coordinate of $AdS_5$. 
This case is the basis of the   ``non-standard'' AdS/CFT correspondence 
suggested in hep-th/0010106.
\\


\end{abstract}

\end{titlepage}
\setcounter{page}{1}
\renewcommand{\thefootnote}{\arabic{footnote}}
\setcounter{footnote}{0}

\def \adss {$AdS_5 \times S^5$\ }
\def \N {{\cal N}}
\def \lc {light-cone\ }
\def \ta { \tau}
\def \s { \sigma }

\def \sg {\sqrt {g }}
\def \te {\theta}
\def \vp {\varphi}
\def \gij {g_{ab}}
\def \xp {x^+}
\def \xm {x^-}
\def \p {\phi}
\def \vt {\theta}
\def \bx {\bar x} \def \a { \alpha}
\def \r {\rho}
\def \fourth {{1 \ov 4}}
\def \DD {{\cal D}}
\def \half {{1 \ov 2}}
\def \inv {^{-1}}
\def \ri {{\rm i}}
\def \D {{\cal D}}
\def \DD {{\rm D}}
\def \vr {\varrho}
 \def \diag {{\rm diag}} \def \td { \tilde }
\def \tta {\td \eta}
\def \cA {{\cal A}}
\def \cB   {{\cal B}}
\def \na {\nabla}

\def \foot {\footnote}

\setcounter{footnote}{0}
\renewcommand{\thefootnote}{\alph{footnote}}

\newsection{Introduction}

Further understanding of string theory -- gauge theory duality 
in its simplest AdS/CFT  case  \ci{mald}  
depends on  progress in solving  string theory  in \adss
with R-R flux.  
This string theory is described by 
 a  well-defined light-cone gauge  Green-Schwarz 
 action \ci{MT3,MTT}, which, in  
  contrast to its  flat space counterpart \ci{GSlc},
 is non-linear. In particular, 
it contains   complicated bosonic $S^5$ sigma model part
as well as  terms  quartic in   fermions.
This complexity is not too  surprising: the dual  $\N=4$  super Yang-Mills 
 theory  should  be non-trivial even in the large $N$ limit, 
 with ``observables''
(dimensions of operators, OPE coefficients, entropy, etc.) being  in general 
complicated  functions of  the `t Hooft coupling 
parameter  $\l= g^2_{\rm YM} N$.\foot{
Below we shall always consider the classical or 
 first-quantized string theory, i.e. assume that $N >> 1$.}

The \lc  \adss string action  is also different 
from   the flat space  GS action in that 
it does not have  manifest 2-d  Lorentz invariance.
This is a generic feature of similar \lc gauge string 
actions in curved space, and is 
 not a problem  of principle, 
in particular 
in the phase space  approach,  
where  the natural starting point is the  \lc  Hamiltonian. 
One  way to address  the  corresponding  quantum problem  is to 
expand the bosonic and fermionic string coordinates and momenta (which
are functions of $\tau$ and  periodic functions of $\sigma$)
in  Fourier modes in $\sigma$,   getting a non-linear 
 space-time supersymmetric 
quantum mechanical 
system with  infinite number of bosonic and fermionic 
degrees of freedom.  
To analyse its spectrum  one may  first to 
restrict consideration  to states that do not depend on $S^5$  directions  and 
try to use numerical methods to deal with
remaining   non-linear terms in the Hamiltonian.

One may hope to simplify the problem  of finding the string spectrum 
by considering some special 
limits of the parameters of the Hamiltonian --
$\a'$ (inverse string tension)
and $R$  (radius of $AdS_5$ and $S^5$). 
In   the general case of 
string in curved space of scale $R$,  
one may consider several  independent limits  when 
 $\a'$ or  $R$  are sent separately 
 to 0 or $\infty$   (under the assumption  of a certain   choice of 
 momenta and coordinates   which are held fixed
in the limit)
   \ci{MTT}.
 For example,  in the 
 particle theory  limit
($\a' \to 0$), the string Hamiltonian 
 reduces to the light-cone Hamiltonian   for a 
superparticle in \adss space \ci{met3}, implying 
that
  the ``massless" or zero-mode  spectrum of the superstring
coincides  with the spectrum of type IIB supergravity
compactified on $S^5$  \ci{MTT}.  
Another  is the  null (tensionless)
 string limit $\a'\to \infty$, 
which,   as  in the flat space  \ci{lin},  
 is obtained by  dropping  terms in the Hamiltonian 
 which contain  derivatives with respect  
to the  spatial world-sheet  coordinate $\sigma$.

In the  special  context 
 of the  AdS/CFT  correspondence,  where 
the (large $N$)   boundary 
conformal gauge  theory  should have    spectrum of dimensions 
depending only on one  parameter $\l$,  
the  dual string theory should also effectively contain only one 
dynamical  parameter -- the effective tension proportional to $\sqrt \l$.  
To hope to establish the  AdS/CFT 
correspondence one should thus  make a special 
 rescaling of  the  radial AdS  string coordinate and its momentum 
so that the  Hamiltonian will depend on   $\a'$ and $R$  
only through  { one}  combination --  $\l= R^4/\a'^2$.\foot{Note that 
(combined also with different boundary 
conditions) this  choice of coordinates 
will not allow one to directly  recover the standard flat space 
 10-d  type IIB superstring  spectrum  in the limit $R\to \infty$, 
which of course  is   not a problem as it 
is not  ``seen'' on the  gauge theory side in any obvious limit.
}

 The limit $\l \to 0$   
has recently attracted some   attention 
due to the conjecture that the free   large $N$  conformal  $\N=4$ SYM 
 should 
be dual to a theory of massless higher spin fields in $AdS_5$
\ci{sund,witt,vv,sez,mih}.
The basic observation is that the free gauge theory has
conserved traceless conformal higher spin currents  \ci{konsh}
 that should  be dual to 
(i.e. couple to the restriction to the boundary of)
 massless higher 
spins in $AdS_5$. 
It would be important of course to see
 the presence of these
massless higher spin states   directly  in 
the spectrum of the \adss string  in the limit $\l \to 0$.

\bigskip

With this  motivation in mind, 
below  we shall supplement 
the general expressions for the \adss  string  action and 
 Hamiltonian  given  in 
 \ci{MTT} with the explicit discussion  of 
 the   $\l \to 0$ limit.
For generality, we shall  consider 
two different  procedures of taking 
 the $R \to 0$ limit, 
  depending on how one scales  the radial coordinate
of the AdS space.
 The first one  -- relevant for the  standard AdS/CFT 
correspondence
 -- starts with   the Hamiltonian
depending on $R$ and $\a'$ only  through $\sqrt \l= R^2/\a'$. Here 
 $R\to 0$ for fixed $\a'$ is  
 equivalent to  the zero tension limit $\a' \to \infty$ for fixed $R$, 
i.e. one effectively keeps the \adss structure of  space-time in the limit. 
Another    $R\to 0$   limit   can be taken  in the action 
chosen   in the form 
which has   straightforward  flat space limit  when  $R \to \infty$.
Here $R \to 0$  means shrinking $S^5$ to zero size and  also 
``freezing'' the radial direction of $AdS_5$. 
Starting with  the string action  written 
in covariant $\kappa$-symmetry gauge this second limit  was considered 
previously in 
 \ci{sina}  in the context of   an alternative  version of the  AdS/CFT 
(\adss string -- $\N=4$ SYM)  correspondence.

\bigskip

Before turning to the discussion    of the string actions
and their  limits 
let us make   few general comments on the ``kinematics'' of the 
AdS/CFT correspondence in the $\l=0$ limit. 
The standard AdS/CFT correspondence  in the sector of  5-d supergravity 
fields or  boundary operators from the stress tensor supermultiplet
can be described  by coupling the  $\N=4$ SYM superconformal 
currents to $\N=4$ conformal supergravity multiplet and then integrating 
over the  $\N=4$ SYM  fields. The resulting large $N$  effective action 
should then reproduce 
 the 5-d gauged supergravity action  
(the quadratic and cubic terms in it, and,  for large $\l$, 
 also higher order terms)  evaluated 
on the solution of the Dirichlet problem \ci{mald}.
In particular, this  was demonstrated explicitly  at the quadratic level 
in the conformal graviton  \ci{liu}.\foot{Coupling the quantum SYM theory 
 to  background 
conformal supergravity multiplet and integrating out the SYM fields 
one finds  the induced  action 
for the conformal supergravity fields:  
$ S =  \int  C_{mnkl}   \ln (\epsilon^2 D^2 )  C_{mnkl} + ...
=             c_1   \int  (  C^2_{mnkl} + ...)  + ${ non-local} terms.
The quadratic and cubic terms in this action expanded in powers of the 
 fields 
 summarize information about protected correlators like 
$< T_{mn} T_{kl}>$ and $< T_{mn} T_{kl} T_{sr}>$   in  conformal
SYM theory.
The same non-local action comes out of solving the  Dirichlet problem 
in the 5-d  $\N=8$ gauged supergravity on the $AdS_5$ background. 
}
  
The same ``induced action'' procedure can be repeated 
for the free $\N=4$ gauge  theory which can be coupled 
to a  
higher spin 
generalization of the conformal supergravity multiplet.
 Coupling  conserved traceless bilinear currents to  the 
corresponding  higher spin conformal  fields, 
integrating out the free SYM fields (i.e. computing the logarithm 
of the determinants in the background)
 and expanding the resulting effective action 
 to quadratic order in the  background
 fields one may then compare 
this quadratic term  to the classical 
 free action of the higher spin fields in $AdS_5$ 
evaluated on the solution of the Dirichlet problem. 
As in the case of the standard 
 conformal  supergravity,  the 
agreement between the two actions is  essentially kinematical, 
i.e. guaranteed  by the symmetries.
Related remarks and discussions of higher spin AdS/CFT correspondence 
appeared in \ci{vv,sez} and in \ci{met2,met4} (in \lc  gauge) and \ci{mih}
(in covariant gauge).\foot{Note that the  $\l=0$ theory 
 will  contain  several currents or conformal  fields 
 of the same spin.  
The corresponding  generalization of conformal supergravity  should  
thus include, in particular, several \ci{sund}  spin 2  conformal gravitons.
 Such theories were recently discussed in  \ci{pvn}.}

More explicitly, 
the quadratic term in the  local (logarithmically divergent) 
part of the corresponding 4-d effective action   will 
be the free conformal higher spin   action 
 which has the following structure
 \ci{us}

\noindent
$
L= \sum_{s=1,2,...}  \p_s \del^{2s} P_s \p_s 
+ \sum_{s=1/2,3/2,...}  \psi_s \del^{2s-1} \g^m \del_m P_s \psi_s . $

\noindent
Here  $s=2$ and $s=3/2$ terms correspond to the standard 
Weyl graviton and gravitino 
of conformal supergravity \ci{PVN}. 
$P_s$ are the transverse traceless projectors defined on totally symmetric
 fields \ci{us}. Let us  consider only the symmetric 
 bosonic fields $\p_s=
(\p_{m_1...m_s})$.\foot{Interacting
 higher spin conformal field  theories  should  exist in flat space 
and have dimensionless cubic, etc., couplings \ci{fl,fll}.
Containing higher derivatives  in the kinetic terms they  
non-unitary (like conformal supergravity or  Weyl gravity), but 
non-unitarity is not an issue in the present 
context.
Here  such  theory is viewed as  an induced (non-local) 
effective field theory   which is simply  a way of 
encoding the information about the  correlators of higher spin 
conformal currents of the SYM theory.
 Refs.  \ci{fl,fll}  speculated about 
possible connections between  conformal higher spins in flat space, 
massless higher spins in AdS, and string theory, but they were not 
anticipating the $4 \to 5$ holographic jump in the dimension essential 
for  the AdS/CFT correspondence. }
The fields $\p_s$ have canonical 
dimension of (length)$^{s-2}$  and can be coupled as $\int d^4 x \ \p_s(x)  J_s(x) $ to 
the traceless currents $J_s= X \del^s P_s X  +...$ 
which are conserved when $X$ is on shell ($X$
with canonical  dimension (length)$^{-1}$
 are the  scalars of the $\N=4$  vector multilet). 
Starting with  $L= X \del^2 X +  \p_s (x)   X \del^{ s} P_s  X  + ... $
and integrating out the SYM fields gives the  induced  action 
$\Gamma[\p_s] \sim \ln \det (- \del^2  +  \p_s    \del^{ s} P_s   ) $.
The  quadratic term in this action  will be proportional to 
$ \int d^4 p\  \p_s (p)   p^{2s} \ln (\ep^2 p^2)  P_s  \p_s(-p) $ or,  
in coordinate representation,   $
 \int  \p_s (x)  \del^{2s} \ln (\ep^2 \del^2)   P_s  \p_s 
$, with the local (divergent) part being  the  above 
pure conformal higher spin action.
This quadratic  term  in $\Gamma[\p_s]$  may be written also as 
$ 
  \int d^4 x d^4 x'    { \p_s (x) \hat  P_s(x-x')   \p_s (x')  \ov (|x-x'|^2 
   + \epsilon^2)^{ s + 2} }  
.$ 
On symmetry grounds, the same  non-local functional  must follow
from the  solution of the Dirichlet problem  for 
the corresponding massless higher spin in $AdS_5$.\foot{Including non-linear in $\p_s$ terms in the coupling to SYM  fields 
will  not change the value of  the conformal 
current correlators at separated points 
but is necessary for maintaining manifest higher spin symmetries 
in the induced action. Let us note in passing 
 that viewing the $\N=4$ conformal supergravity action as the local part of the induced action generated by integrating out 
the $\N=4$ SYM fields in the   conformal supergravity  background  
 implies that 
the conformal supergravity action must contain the Weyl
 tensor term $C^2_{mnkL}$
without any  extra scalar-field dependent  prefactor, cf. \ci{FT,us}.}

\section{$R\to 0$ limit in string action  in covariant gauge}	

We shall start with  a  heuristic 
 discussion of the $R \to 0$ limit  taken 
directly in the \adss superstring  action in   covariant gauge. 

 The 
 superstring Lagrangian \ci{MT1}  may be 
  written in  ``4+6" parametrization 
 of \adss  corresponding to the metric  ($a=0,...,3, \ 
 M=1,...,6$, $Y^M=  e^\p u^M$,  $u^M u^M =1$)   
 \be
 ds^2 = Y^2 dx^a dx^a + R^2 Y^{-2} dY^M dY^M  
=  e^{2 \p} dx^a dx^a  + R^2  d\p^2 +  R^2  du^M du^M \ . \la{mmm}
 \ee
This metric has the standard flat space limit  when  $R\to \infty$
(setting $\p= \vp/R$ but without need to rescale $x^a$). 
Another  parametrization  is based on rescaling $Y$ by $R$ 
(i.e. shifting $\p$ by $\ln R$). Then  $R^2$ is  the factor 
in front of the whole 10-d metric 
\be
 ds^2 = R^2 ( Y^2 dx^a dx^a + Y^{-2} dY^M dY^M ) 
= R^2 ( e^{2 \p} dx^a dx^a  +   d\p^2 +   du^M du^M) \ . \la{mmme}
\ee
This second choice  is the basis for the standard AdS/CFT 
correspondence where the string action will then depend on $R$ 
through $\sqrt \l = R^2/\a'$. 
This choice means  that  all distances, including 
the world-sheet ones are  effectively  measured in terms of $R$. 

Interpreting  the   \adss isometry  supergroup  $PSU(2,2|4)$ as 
 the   $\N=4$ 
 superconformal group in 4 dimensions, it is natural to split 
 the  fermionic generators into 4 standard supergenerators $Q_i$
 and  4 special conformal supergenerators $S_i$ (we suppress the  
 4-d spinor indices). The associated 
 superstring coordinates will be denoted as 
 $\theta_i$ and $\eta_i$ 
 \ci{MT3,MTT}.
 The 4-d Lorentz covariant $\k$-symmetry 
  ``S-gauge" of \ci{MT3}
 corresponds to setting all $\eta_i$  fields  to zero.
The resulting Lagrangian has the following simple structure
 \ci{MT3}\foot{The actions in 
 \ci{Pes,KR,sira} have  equivalent  forms, corresponding to  particular 
 choices
 of the 10-d Dirac matrix representation.
We use the following notation:
the 4-d indices are $a,b=0,1,2,3$; $SO(6)$ indices are 
 $ M,N=1,...,6$; $SU(4)$ indices are $i,j=1,2,3,4$;
the 2-d indices are 
$\mu,\nu=0$.  
 We use  ``chiral" representations for the 4-d and 6-d Dirac matrices,
 $\g^a =\left(\begin{array}{cc}
 0   & \s^a
 \\
\bar  \s^a & 0
 \end{array}
 \right)$, \ 
 $ \gamma^M
=\left(\begin{array}{cc}
 0   & \rho^{Mij}
 \\
 \rho^{M}_{ij} & 0
 \end{array}
 \right)\,, $  with 
 $(\rho^M)^{ij}\equiv  - (\rho_{ij}^{M})^*$.
 $\sfa,\dsfb=1,2$ are   the  $sl(2,C)$ 
 (i.e. 4-d spinor) indices and 
 $\theta_{\sfa i}^\dagger = -\theta_{\dsfa}^i$, 
 $\theta^{\sfa}_i{}^\dagger = \theta^{\dsfa i}$. 
 The 10-d spinors are split as $(\t^i,\t_i)$.}
 \be
 L= - \frac{1}{2}\sqrt g 
 [ Y^2 (\del_\vm x^a
  - {\rm i} \theta  \sigma^{a }\del_\vm \theta )^2 
 + R^2 Y^{-2}\del^\vm Y^M \del_\vm Y^M ]  
  -  {\rm i} \epsilon^{\mu\nu} R \del_\mu Y^M 
   \theta \rho^M \del_\nu  \theta \  . \la{bss}
   \ee
Here
$$ {\rm i} \theta  \sigma^{a }\del_\vm \theta
\equiv 
{\rm i} \theta_{\sfa i} 
  \sigma^{a\sfa\dsfb}\del_\vm \theta_{\dsfb}^i 
     + h.c. \ , \ \ \ \ \ 
{\rm i}  \theta \rho^M \del_\nu  \theta
\equiv 
 {\rm i}  \theta^{\dsfa i} \rho^M_{ij} \del_\nu  \theta_{\dsfa}^j +
   h.c. \ . 
$$
We have written the Lagrangian corresponding 
to the metric \rf{mmm}, where  $Y^M$ is dimensionless 
($x^a$  and $R$ have dimension of length 
  and  the  fermionic coordinates $\theta$ 
  have dimension (length)$^{1/2}$). 
 The resulting  string 
 action $I = { 1 \ov 2\pi \a'} \int d^2 \s  \ L$ is then 
 dimensionless.  To see that the  action \rf{bss} 
 has the correct flat space limit one is to set 
 $\phi = \vp/R$  and send $R$ to infinity.

The action  corresponding to \rf{mmme} is obtained by $Y^M \to R Y^M$:
 \be
 L= R^2 \bigg\{ - \frac{1}{2}  \sqrt g 
 [ Y^2 (\del_\vm x^a
  - {\rm i} \theta  \sigma^{a }\del_\vm \theta )^2 
 + Y^{-2}\del^\vm Y^M \del_\vm Y^M ]  
  -  {\rm i} \epsilon^{\mu\nu}  \del_\mu Y^M 
   \theta \rho^M \del_\nu  \theta \bigg\} \  . \la{bsse}
   \ee
As discussed in the Introduction, the 
 way to take the   $R \to 0$ limit is in general 
 not  unique: 
the result depends on additional assumptions about which coordinates 
and arguments  (momenta) of external states are held fixed in the limit. 
For example, starting with   \rf{bss} or \rf{bsse}
gives apparently different results.
The action corresponding to  \rf{bsse} 
depends on $R$ through 
 the dimensionless ratio $\l^{1/4} = R/\sqrt{\a'}$, so that 
$R \to 0$  corresponds  to $\l\to 0$ and thus is also equivalent to
$\a' \to \infty$  with  $R$ 
(and external momenta)   kept finite in this  limit.

In the rest of this  section we shall follow  \ci{sina}  and  discuss 
taking the $R \to 0$ limit in  
 the case of \rf{bss}. The direct  result of  setting  $R=0$ in 
  \rf{bss} is 
 \be \la{limi}
 L_{R\to 0} = - \frac{1}{2}\sqrt g \  Y^2\  (\del_\vm x^a
  - {\rm i} \theta
  \sigma^{a}\del_\vm \theta )^2  \ , 
 \ee
suggesting  that  six  $Y^M$  coordinates effectively ``decouple''. 
 This truncated action  still  has the expected $\N=4$  global linear
supersymmetry and 4-d conformal invariance 
 ($Y$ is still to be integrated over).

The  action \rf{limi} does not of course define a consistent critical string 
theory and should be  supplemented with  a more systematic 
procedure  of taking the $R\to 0$ limit in  
the path integral with external operator insertions. 
 Since the coefficient of the kinetic term of $Y^M$ directions
 goes to zero in the limit, that means   
 that if  the external operator 
 depends on {  non-constant} modes of $Y^M$  then the $R\to 0$ 
limit  will be  singular. 
To  be able to define this limit, 
one should thus restrict consideration 
to  external states that do not depend on non-constant modes in 
radial $AdS_5$  and all  $S^5$ 
  directions. 
However, 
one should still integrate over the non-constant  modes 
of $Y^M$ in the path integral,
and this is 
essential to ensure that 
the resulting string theory  is consistent.
Indeed, 
 the \adss string sigma model \rf{bss} is conformal and has the 
 right central
 charge for any $R$ \ci{MT1,KKK}, so 
 taking the limit after doing the integral over 
all of the fields including 
the  non-constant modes of $Y^M$   
should   not affect these properties. 

 The constant  modes  of $Y^M$  
should  in general  couple to  external states 
 and the integral over them will   be non-trivial, i.e. we should get, symbolically, 
\foot{Again, it is understood that 
   one has  first  integrated out all non-zero modes of $Y^M$
  to cancel the conformal anomaly  and ensure that the model is finite.}
   \be 
   <F>= \int  d^6 Y_0 \ M(Y_0)   \ \int D x  D\theta \ 
   e^{ {\rm i} I [  x,\theta,  Y_0] } \  F[ x ,\theta,  Y_0]  
 \ , \ee 
 where $Y_0$ is the constant part of $Y$.
 We have added the measure factor $M(Y_0)= \sqrt {G_{10}} = Y^{-2}$ 
(but  did 
not explicitly split  $x$ and $\theta$  on  constant and non-constant modes).
 Assuming further that external states do not depend on $S^5$ angles
(i.e. $F$  is $SO(6)$ symmetric),  we get 
 (setting $¦Y_0¦= e^{\p_0}$, $T= {1 \ov 2 \pi \a'} $)
 \be \la{plo}
  <F> \sim   \int  d \p_0 \  e^{4 \p_0}   \ \int D x \ D\theta\ 
   e^{ -  \frac{i}{2} T \int d^2 \s
   \sqrt g \  e^{2 \p_0}  (\del_\vm x^a
  - {\rm i} \theta \sigma^{a}\del_\vm \theta)^2  } 
     \  F[x ,\theta,  \p_0]  \ . 
 \ee
 The integral over the constant  $\po$  should then  ensure the target space 
scale invariance in 4 dimensions, i.e. the symmetry 
  under rescalings of $x^a$.

To summarize, in the particular $R\to 0$ limit in \rf{bss} 
corresponding to \rf{mmm}  the string is effectively allowed
to fluctuate only in 4  space-time dimensions (as well as in  
 fermionic dimensions),
and the  integration over the constant mode $\p_0$
of the radial $AdS_5$ direction 
 produces  averaging over 
 the effective string tension  ($\TT \equiv  T e^{2 \p_0}$)
 \be \la{avv}
  <F> \sim  \int_0^\infty   d \TT \  \TT^{3/2}    \ \int D x \ D\theta\ 
   e^{ - \frac{i}{2} \TT  \int d^2 \s
   \sqrt g \   (\del_\vm x^a
  - {\rm i} \theta \sigma^{a}\del_\vm \theta)^2  } 
     \  F[x ,\theta, \TT]  \ . 
 \ee
If instead one starts with the action \rf{bsse}  
then the $R \to 0$   limit   
 is equivalent to the zero tension limit  $T \to 0$ 
  and is apparently a   strong-coupling limit in 
all 10 directions (see also  \ci{poll} for related remarks). 
 In general, the covariant gauge action is  not 
the best  starting point for the investigation of the $R\to 0$ limit
 because it  
 does not  allow one 
 to find the spectrum of ``small'' strings (the  kinetic term of 
 fermions is not well defined). 
 We shall therefore turn  now 
to the discussion  of  the  $R\to 0$  limit in the \lc gauge.

\section{Superstring action in fermionic  \lc  gauge}
\noindent

 In flat space, the  superstring  \lc gauge  fixing 
includes the  fermionic \lc gauge choice
(i.e. fixing the $\k$-symmetry by  the $\g^+ \theta^I=0$ condition), 
 and the  bosonic \lc gauge choice  (i.e.
using   the conformal gauge\foot{Below we  shall  use the Minkowski
signature 2-d world sheet  metric $g_{\vm\vn}$ with
$g\equiv - \det g_{\vm\vn}$.}
$\sqrt {g}  g^{\vm\vn} =\eta^{\vm\vn}$
and fixing  the  residual conformal diffeomorphism symmetry
by $x^+(\ta,\s)  =  p^+ \tau$).
Fixing the  fermionic \lc gauge
  produces already  a substantial
simplification of the flat-space
GS action: it   becomes  quadratic in $\theta$.
Similarly, in \adss one is able to  
  find a light-cone 
  $\k$-symmetry gauge \ci{MT3}
  in  which the fermion kinetic term 
  has the structure  $  \del x^+ \theta  \del \theta $. It thus 
    involves only one
  combination -- $x^+$ -- of the 4-d coordinates, so that  
 the non-degeneracy of the fermion  kinetic term
does  not
 depend on a choice of  specific  string background
  in transverse directions.
  To simplify the fermion kinetic term further 
  one may    choose next  the   
  bosonic \lc gauge $x^+ = p^+  \tau$. Fixing the bosonic part of the 
\lc gauge 
and derivation of the resulting \lc Hamiltonian 
was described  in detail  in \ci{MTT} (see also \ci{TS})
and will be discussed in the next section.

Let us first review the form of the string action found after fixing 
the \lc $\kappa$-symmetry gauge. 
In contrast to the Lorentz covariant ``S-gauge" where 16 fermions 
 $\eta_i$ are set equal to zero, the light-cone 
 gauge  used in  \ci{MT3} corresponds to setting to zero 
 ``half" of the 16 $\theta_i$ and ``half" of the 16 $\eta_i$
fermionic coordinates
 (``half" is defined with respect to $SO(1,1)$ rotations in the 
 light-cone directions). The remaining fermions
(which will be again   denoted  by  $\theta_i, \eta_i$)
from  now on  will be   simply 4+4  complex anticommuting variables 
 carrying {no } extra  Lorentz spinor  indices.
 For comparison,  the flat space  GS action in the 
 light-cone gauge ($ \gamma^+ \theta=0$) 
  \ci{GSlc}, written in a  similar parametrization 
 of the 16 fermionic
 coordinates, has the following structure
 (suppressing the $SU(4)$ indices)
  \be\la{fla}
{ L}
=- \frac{1}{2}\sqrt{g} (\partial_\vm x)^2 - \Bigl[
 \frac{{\rm i}}{2}\sqrt{g} \partial^\vm x^+(
\theta\partial_\vm \theta
+\eta\partial_\vm\eta)  - \epsilon^{\vm\vn}
\partial_\vm x^+ \eta\partial_\vn\theta+h.c.
\Bigl]\ .\ee 
This form of the  original  GS Lagrangian  is indeed  the 
flat space ($R\to \infty$) limit
 of  the  \lc  \adss  Lagrangian   of \ci{MT3}
written in the parametrization used in \rf{mmm} (cf. \rf{bss})
$$
{ L} =
- \sqrt{g}\Bigl[
Y^2(\partial^\vm x^+ \partial_\vm x ^-
+ \partial^\vm x\partial_\vm\bar{x})
+\frac{1}{2} R^2 Y^{-2} ( \del_\mu Y^M +  
{\rm i} R^{-2}  \eta\rho^{MN}\eta  Y^N Y^2 \del_\mu x^+ )^2      \Bigl]
$$
$$ -  \ \bigg(  \frac{{\rm i}}{2} \sqrt{g}g^{\vm\vn}
Y^2\partial_\vm x^+
\big[\theta^i\partial_\vn \theta_i
+\eta^i\partial_\vn \eta_i
+{ {\rm i}\ov 2} R^{-2}  Y^2\partial_\vn x^+(\eta^2)^2\big] 
$$ \be
- \ \epsilon^{\vm\vn}
|Y|\partial_\vm x^+ \   \eta^i \rho_{ij}^M Y^M
(\partial_\vn\theta^j-{\rm i}R^{-1} \sqrt{2}|Y| \eta^j
\partial_\vn x)+h.c. \bigg)  \ . \label{actki}
\ee
$x^a$ are decomposed into the light-cone $(x^+,x^-)$ and  2 complex transverse 
coordinates
$x_\perp=(x, \bx)$,  i.e. $
x^\pm= \frac{1}{\sqrt{2}}(x^3\pm x^0)\,; $ $
x,\bx = { 1 \ov \sqrt 2} (x^1 \pm  {\rm i} x^2)$.\foot{The matrices $\rho^M$ (off-diagonal blocks of 6-d Dirac matrices
in chiral representation) satisfy:
$\rho_{ij}^M =- \rho_{ji}^M\,,$ $ 
   (\rho^M)^{il}\rho_{lj}^N + (\rho^N)^{il}\rho_{lj}^M
 =2\delta^{MN}\delta_j^i\,, $ $
  (\rho^M)^{ij}\equiv  - (\rho_{ij}^{M})^* ,$
  and $\rho^{MN} \equiv \rho^{[M} \rho^{\dagger N]} $, i.e. 
  $\rho^{MNi}_{\ \ \ j} =\frac{1}{2}(\rho^M)^{il}\rho_{lj}^N
-(M\leftrightarrow N)\,.$ 
Also, $ \theta_i^\dagger =\theta^i\,,$ $
\eta_i^\dagger = \eta^i\ , $
$
\theta^2 \equiv \theta^i\theta_i\,, \ 
\eta^2 \equiv \eta^i\eta_i\,.$ } 
The superconformal algebra $psu(2,2|4)$  implies  that
the   16 physical Grassmann variables --
``linear"  $\te^i$  and   ``nonlinear" $\eta^i$
and their hermitian conjugates $\te_i$  and $\eta_i$
which  transform
in the  fundamental representations of $SU(4)$ 
 are  related to  the  Poincar\'e
and  the  conformal
supersymmetry   in the \lc gauge description of the
    boundary theory.\foot{The action is invariant  under shifting $\theta\rightarrow
\theta+\epsilon$
  (supplemented with a shift  of $x^-$).}

The dependence on the radius $R$
is consistent with
$\theta$ and $\eta$  having 
dimensions (length)$^{1/2}$  and    $Y^M$ being dimensionless in \rf{mmm}.
The $\eta$-fermionic terms with $R^{-2}$  factors correspond to 
the connection (and R-R coupling) 
 and curvature terms that go away in the flat space limit
($|Y|\to 1$) where one recovers \rf{fla}.\foot{The presence of the  $\eta^4$ term   reflects
the curvature of the background; 
the  `extra' $O(\eta^2)$ terms  have the interpretation of the
couplings to the R-R  5-form  background \ci{MT1}.}

Written in terms of the radial direction $\p$ of $AdS_5$ 
and unit 6-d  vector $u^M$  parametrizing $S^5$ the
Lagrangian  \rf{actki} becomes ($Y^M= e^\p u^M$, $|Y|= e^\p$) \ci{MT3}
$$
{L} =
-\sqrt{g}
e^{2 \p} (\partial^\vm x^+ \partial_\vm x ^-
+ \partial^\vm x\partial_\vm\bar{x}) $$
$$- \frac{1}{2}  \sqrt{g} R^2  \big[   \del^\mu \p
\del_\mu \p 
+ ( \del_\mu u^M +  
{\rm i} R^{-2} 
\eta\rho^{MN}\eta  u^N  e^{2\p}  \del_\mu x^+ )^2 \big]
$$
$$ - \ \bigg(  \ \frac{{\rm i}}{2} \sqrt{g}g^{\vm\vn}
 e^{2\p} \partial_\vm x^+
[\theta^i\partial_\vn \theta_i
+\eta^i\partial_\vn \eta_i
+{{\rm i}\ov 2}  R^{-2}  e^{2\p} \partial_\vn x^+(\eta^2)^2] 
$$ \be
- \ \epsilon^{\vm\vn}
e^{2\p} \partial_\vm x^+\  \eta^i \rho_{ij}^M u^M
(\partial_\vn\theta^j-{\rm i} R^{-1} \sqrt{2}e^{\p}  \eta^j
\partial_\vn x)+h.c. \bigg) \ . \label{tki}
\ee
Starting  with the action \rf{tki},  
to get a regular  $R\to 0$ limit   we  need to  deal with blowing up
of the fermionic  terms with $R^{-1}$
  factors.  
One option is to demand that external states have no dependence on $\eta$, i.e.
to   effectively  freeze  $\eta$ to zero.\foot{A more systematic procedure 
is to  introduce  $\z \equiv   R^{-1} \eta $ 
 which will be fixed in the $R\to 0$ limit.
  Then the $\zeta$-dependence 
 decouples in the $R\to 0$ limit, i.e. this limit is regular provided the external states do not depend on $\z$.} 
By formally setting $R=0$ in  the action 
 we are then left with
  \be 
{L}_{R \to 0}  =
-\sqrt{g}
e^{2 \p} \bigg[ \partial^\vm x^+ \partial_\vm x ^-
+ \partial^\vm x\partial_\vm\bar{x}
  +  \ \partial^\vm x^+ 
\big(  \frac{{\rm i}}{2}  \theta^i\partial_\vm \theta_i
+h.c. \big) \bigg] \ . \label{limm}
\ee
  This is the \lc  gauge analog of 
  the  ``naive'' action  \rf{limi}. 
 As in the previous section,  to 
 get a well-defined  limit we should again average only  quantities
  that do not have  dependence on {non}-constant modes of 
  $\phi$, $u^M$ (which should be  integrated out before taking
 $R\to 0$). 
Since the integral over  $\p$  will be  effectively restricted to 
constant values
 it is not problem to fix the standard 
  bosonic  light-cone gauge  $x^+ = p^+ \tau$.  Note that in 
 contrast to the standard flat space GS action the
 resulting fermionic term in the action will miss the spatial 
derivative part, i.e. will not be 2-d Lorentz covariant.
As discussed in \ci{MT3,MTT},  this is a general feature of the \lc gauge 
actions in curved AdS-type spaces. 

  
Let us now turn to the case of the action  which is relevant 
for the standard AdS/CFT correspondence, i.e. the one associated  
the metric \rf{mmme} and  thus related to \rf{tki} by 
 $e^\p \to R e^\p$, 
$$
{L} = R^2 \bigg\{
-\sqrt{g}
e^{2 \p} (\partial^\vm x^+ \partial_\vm x ^-
+ \partial^\vm x\partial_\vm\bar{x}) 
 - \  \frac{1}{2} \sqrt{g} \big[    \del^\mu \p
\del_\mu \p 
+  ( \del_\mu u^M +  
{\rm i}  
\eta\rho^{MN}\eta  u^N  e^{2\p}  \del_\mu x^+ )^2 \big]
$$
$$ - \ \bigg(  \ \frac{{\rm i}}{2} \sqrt{g}g^{\vm\vn}
 e^{2\p} \partial_\vm x^+
[\theta^i\partial_\vn \theta_i
+\eta^i\partial_\vn \eta_i
+{{\rm i}\ov 2}    e^{2\p} \partial_\vn x^+(\eta^2)^2] 
$$ \be
- \ \epsilon^{\vm\vn}
e^{2\p} \partial_\vm x^+\  \eta^i \rho_{ij}^M u^M
(\partial_\vn\theta^j-{\rm i}  \sqrt{2}e^{\p}  \eta^j
\partial_\vn x)+h.c. \bigg) \bigg\}\ . \label{atki}
\ee
As in the covariant gauge action    \rf{bsse} of  the previous section, 
here    $R\to 0 $   is  the 
strong-coupling  limit in all directions,  and there  seems to be 
 no natural 
 simplification of the action.  
In  such  cases it is more appropriate to use the 
(\lc gauge) phase space formulation
to the discussion of  which we now turn.

\section{Bosonic \lc  gauge fixing  in $AdS$  space 
}
\noindent

To eliminate the $\del x^+ $-factors from the fermion kinetic
 terms  in \rf{tki} or \rf{atki}  one is to fix the bosonic \lc gauge.
 In flat space this 
may be  done by fixing the conformal gauge
   $
   \sg g^{\vm\vn}=\eta^{\mu\nu} ,$
     and then noting
   that 
   one can fix the residual conformal diffeomorphism symmetry
   on the plane by choosing $x^+(\tau,\sigma) = p^+  \tau$.
In the case of the  AdS-type curved spaces, 
the bosonic \lc gauge $x^+=p^+ \tau$   
can not be combined with the standard conformal gauge
$\sqrt g g^{\m\nu}= \eta^{\mu\nu}$.
One needs  instead to impose a condition
on $g_{\m\nu}$ that breaks the manifest  2-d Lorentz symmetry.
One  consistent gauge choice  is \ci{MT3}:  
 $
 e^{2\phi}\sqrt g g^{00}=-1 , 
 \ x^+ =  p^+ \tau  .$ 
Using these conditions in the action \rf{tki} and integrating over 
$x^-$ and the remaining components of the 
2-d metric  one finds the explicit form 
of the action which is equivalent to the one obtained in 
\ci{MTT}  in the phase space approach. 
A closely related   alternative originally suggested in  \ci{pol}
is a  modification of the conformal gauge 
$\sqrt g g^{ab} = \diag(- e^{-2\p}, e^{2\p})$ 
which, as in flat space,  it  turns out to be possible to 
supplement  with  the third  \lc gauge condition 
$x^+=p^+ \tau$.\foot{Equivalent action is found also using the gauge \ci{PS}
$g_{01}=0, \ x^+=p^+ \tau$.} 
   
The resulting action has $AdS_5$ radial direction $e^\p$
 factors coupled differently to $\del_0$ and $\del_1$
 derivative terms, and   the $S^5$ part of
  the action is also coupled to $\p$. 
In the   absence  of manifest 2-d Lorentz symmetry   
it is natural to use the {phase space}  formulation   of 
the \lc gauge fixed  theory. 
The coordinate space Polyakov 
   approach based on   any of the  above gauges    is equivalent
 to the phase space  GGRT approach 
   based on  fixing the diffeomorphisms by 
   $x^+ =  \tau, \ P^+  =p^+$=const \ci{MTT}. 


To illustrate the derivation of the  phase space Lagrangian  
let us  first ignore the  $S^5$ and 
fermionic  parts in \rf{atki}, i.e. start with  
\be
{\cal L} \equiv T L  = -   \T  h^{\vm\vn} \big(  
\partial_\vm x^+ \partial_\vn x^-
+ \partial_\vm x\partial_\vn\bar{x}
 + \ha  e^{-2\p}\del_\vm \p\del_\vn \p\big) \ ,  \la{lll}
 \ee
\be \la{tenn} 
 \T \equiv T R^2  = { R^2 \ov 2 \pi \a'} = { \sqrt \l \ov 2 \pi} 
 \ , \ \ \ \ \ \ \ \ \ \ \ \  h^{\vm\vn} \equiv \sqrt g g^{\vm\vn} e^{2\p} \  .  \ee 
Introducing the momenta $\PP^a= \T { \del L \ov \del \dot x^a} $ for the \lc coordinates $(x^+,x^-)$
and the two transverse coordinates $x_\perp=(x,\bar x)$ and the momentum $\Pi$ for 
the radial direction $\p$ we get 
$$
{\cal L} = \dot x_\al  \PP_\al   + \dot \p \Pi + \dot x^+ \PP^-  +
\dot x^- \PP^+  +  { 1 \ov 2 h^{00}} 
\bigg[ \T\inv  \PP^2_\al + \T\inv   \PP^+\PP^- 
$$
\be + \     \T e^{4\p}   (  \x'^2_\al  +  2\x'^+ \x'^-) + e^{2\p} 
(\T\inv  \Pi^2 +  \T   \ph'^2 ) 
\bigg]
\la{logj}  +  { h^{01} \ov  h^{00} }
( \x'_\al \PP_\al 
+ \ph'\Pi  + \x'^+ \PP^-  + \x'^- \PP^+) \ , 
\ee
where $ 1/h^{00}$ and  ${ h^{01}/h^{00} }$ play the role of the
Lagrange
multipliers imposing two constraints. 
Choosing the \lc gauge  $
  x^+ = \tau\ , \  \PP^+  = p^+  = \const $  and 
integrating over $\PP^-$ we get the relation 
$
h^{00}    = - \T^{-1} p^+  . $
The expression for $\PP^-$ follows from the $1/h^{00}$
constraint after using  the $h^{01}/h^{00}$  constraint.
The resulting \lc gauge Hamiltonian density ${\cal H}$
 is  \ci{MTT}   
\be
{\cal H} =-\PP^- = {1 \ov 2 p^+ } 
  \bigg[  \PP^2_\al  + e^{4\p} \T^2  \x'^2_\al 
 +  e^{2\p}( \Pi^2 +  \T^2  \ph'^2)  \bigg] \ . 
\la{hamm}
\ee
The constraint determining $x^-$ is 
$ 
p^+\x'^-  + \PP_\al\x'_\al   + \ph'\Pi =0 $.
The    coordinates and momenta satisfy
the usual Poisson bracket relation
$
\{ x(\tau, \sigma), \PP(\tau,\sigma')\} = \delta( \sigma - \sigma') .
$
Integrating over the momenta $ \PP_\al$ and $\Pi$, one finds the corresponding 
\lc gauge Lagrangian 
\be
L =  {p^+ \ov 2 } 
  \bigg( \dot x^2_\al  -  e^{4\p} \T^2  (p^+)^{-2} \x'^2_\al 
 +  e^{-2\p}[ \dot \p^2 -   e^{4\p} \T^2 (p^+)^{-2} \p'^2]  \bigg) \ .  
\la{lamm}
\ee 
As usual, the  explicit dependence on $p^+$  can be absorbed into redefinition 
$\s \to \s/p^+$ in the  action. 
The analog of \rf{hamm}  in the case of \rf{mmm}, i.e. the 
action \rf{tki},  is 
obtained by 
 $e^\p \to R^{-1} e^\p$, 
\be
{\cal H} =-\PP^- = {1 \ov 2 p^+ } 
  \bigg[  \PP^2_\al  + e^{4\p} T^2  \x'^2_\al 
 +  R^{-2} e^{2\p}(  \Pi^2 +  T^2 R^4  \ph'^2)  \bigg] \ . 
\la{hmm}
\ee
In  this case    $\cal H$ depends separately on 
the tension $T$ and the radius $R$, and so  one may 
 consider  several  limits 
(assuming other arguments of $\cal H$  are kept fixed): 
(i) $T\to \infty, \ R=$fixed: particle in AdS;
(ii) $T\to 0, \ R=$fixed: ``standard'' tensionless string  in AdS;
(iii) $R\to \infty, \ T=$fixed: flat space;
(iv) $R\to 0, \ T=$fixed: string in  ``zero radius''  AdS. 
To get   finite  result   in the particle  limit  
 $T\to \infty$ 
 one  must set $x'=0, \ \p'=0$, 
i.e. assume that the  string   shrinks to a point,  
$
{H}_{T \to \infty}   = {1 \ov 2 } 
  \big( \PP^2_\al   
 +  e^{2\p}  \Pi^2   \big)  . $
The  limit $R\to \infty$ in  \rf{hmm}  is defined by setting 
$\p = \vp/R, \ \Pi = R p$; this preserves the Poisson bracket 
and gives the standard  regular  flat space expression.
To define the limit $R\to 0$ in  
  \rf{hmm}\  we must (as in the previous section)
restrict consideration to  states that { do not  depend on  } $\Pi$, 
i.e.  have zero momentum in the radial AdS direction,   
$
{\cal H}_{R \to 0}   = {1 \ov 2 p^+ } 
  \big( \PP^2_\al  + T^2  e^{4\p}  \x'^2_\al  \big)$.

In  the case of \rf{hamm}  the flat space limit is not reached directly, and 
there are only two independent limits:
$\T \to 0$ and $\T\to \infty$.
Here  the limit 
 $R \to 0$  for fixed $T$   is equivalent to  $\T \sim \sqrt \l\to 0$  and thus to 
 $T \to 0$  for fixed $R$. It thus  corresponds 
(as the usual zero-tension limit in flat space) to 
dropping $\s$-derivative terms in the action and Hamiltonian, i.e. 
\be
{\cal H}_{\l \to 0}  = {1 \ov 2 p^+ } 
  \big(  \PP^2_\al  +   e^{2 \p}  \Pi^2  \big) \ .  
\la{hmmaa}
\ee


\section{Superstring light cone  action  and its $\l \to 0$ limit }

Let us now add the dependence on $S^5$ and fermionic directions.
Repeating the above bosonic \lc 
 gauge fixing  procedure for the superstring action
in \rf{atki} we get \ci{MTT}
$$
{\cal L} 
= 
\PP_\al\dot{x}_\al +  \Pi\dot{\phi} 
+ \PP_M \dot{u}^M
+ \frac{{\rm i}}{2}  p^+(\theta^i \dot{\theta}_i
+\eta^i\dot{\eta}_i   -  h.c. ) - {\cal H}
$$
\be - \ 
T (p^+)^{-1} {h^{01}}\Bigl[p^+\x'^-  + \PP_\al\x'_\al + \Pi\ph'
+ \PP_M \u'^M
+\frac{{\rm i}}{2}p^+(\theta^i\th'_i+\eta^i\et'_i
- h.c. )\Bigr] \  , \label{lag6}
\ee
where  the \lc Hamiltonian density is
$$
{\cal H} = -\PP^-= 
\frac{1}{2p^+}\Bigl(   \PP_\al^2+ \T^2  e^{4\phi}\x'_\al^2 
 +\  e^{2\phi}[   \Pi^2+ \T^2  \ph'^2  
+   \PP^M\PP^M + \T^2  \u'^M \u'^M $$  \be 
+ \   p^{+2}(\eta^2)^2  + \  2   {\rm i} 
  p^+ \eta_i  (\rho^{MN})^i{}_j  \eta^j  u^M \PP^N 
] 
\Bigr) 
-\ 
 \T e^{2\phi} u^M  [ \eta^i \rho^M_{ij} 
(\th'^j - {\rm i}\sqrt{2}  e^\phi \eta^j\x')
\    +   h.c. ] \  .\la{dur}
\ee
Here  the odd 
part of the  phase space  is
 represented by 
 $\theta^i$, $\eta^i$ considered as fermionic 
coordinates and $\theta_i,$ $\eta_i$ considered as 
fermionic momenta. $h^{01}$ imposes the reparametrization constraint,  
and  $\PP^M$ is the momentum  corresponding to the unit 6-vector
$u^M$, \  $u^M \PP^M =0$.
  
In the particle theory limit, i.e. $\T \to \infty, $  
 implying that all spatial derivatives are to be set to zero,
 \rf{dur} reduces \ci{MTT}  to the 
(classical limit of) quantum \lc 
Hamiltonian for a superparticle in \adss space 
\ci{met3}. 
\foot{As a result,  the ``massless" (zero-mode)  spectrum of the superstring
coincides indeed with the spectrum of type IIB supergravity
compactified on $S^5$. The  vertex operators for the 
supergravity states are then  obtained by solving the
linearized 
supergravity equations (or 1-st quantized superparticle state
equations) in \adss background. }  


The equivalent  expressions  for the string Lagrangian and Hamiltonian
written in terms  of 6 Cartesian coordinates 
$Y^M$ and the associated (now unconstrained) 
momenta   $\PP_M$, i.e. 
corresponding to the Lagrangian \rf{actki} but with 
the ``standard AdS/CFT''  rescaling 
$Y^M  \to R Y^M , \ \PP_M \to R^{-1} \PP_M   $
are 
\be
{\cal L} 
= \PP_\al\dot{x}_\al 
+ \PP_M \dot{Y}^M
+ \frac{{\rm i}}{2}p^+(\theta^i \dot{\theta}_i
+\eta^i\dot{\eta}_i-  h.c.)  -
{\cal H}  \  , 
\la{laah}
\ee
$$
{\cal H}  =  
\frac{1}{2p^+}\Bigl( \PP_\al^2 +  \T^2   Y^4\x'_\al^2 + 
   Y^4\PP_M \PP_M  + \T^2   \Y'^M\Y'^M $$ 
$$
+\        Y^2[p^{+2}(\eta^2)^2 + 2{\rm i}p^+\eta \rho^{MN}\eta  Y_M
\PP_N]\Bigr)
 $$
  \be \la{haha}
-\ \T  |Y| Y^M [\ \eta \rho^M 
(\th' - {\rm i}  \sqrt{2}|Y| \eta\x') + h.c.] \  .
\ee
Here also $
p^+\x'^-  + \PP_\al\x'_\al 
+ \PP_M \Y'^M
+\frac{{\rm i}}{2}p^+(
\theta^i\th'_i+\eta^i\et'_i - h.c. )  =0 .$ 


The $\l\to 0$  limit in \rf{haha} is the same as  $T\to 0$, 
i.e.  we get 
\be 
{\cal H}_{\l \to 0}   =  
\frac{1}{2p^+}\Bigl( \PP_\al^2 +   Y^4\PP_M \PP_M    
+\       Y^2[p^{+2}(\eta^2)^2 + 2{\rm i}p^+\eta \rho^{MN}\eta  Y_M
\PP_N]\Bigr)
   \la{hahao} \ . \ee 
In contrast to the tensionless string limit in  flat space 
where one gets an  infinite set of  ``decoupled'' modes, 
here the $S^5$ and fermionic  terms  introduce  non-trivial
interactions between the  Fourier modes in $\s$. 
Note that the \adss 
superparticle   is  effectively the  special case of this limit:
 while sending  $T\to \infty$
implies that $\s$-derivatives are constrained to be 
zero,  their contribution is dropped in the $T\to 0$ limit.


\bigskip

Solving for the   momenta in  the phase space  
action  \rf{laah} one finds the  \lc Lagrangian  corresponding to the 
Hamiltonian \rf{haha}, 
which  generalizes \rf{lamm}. It is useful to present  it 
in terms of  the ``conformally-flat'' 10-d  coordinates $(x^a,Z^M)$,
where $Z^M$ is the ``inverse'' of $Y^M$. The   corresponding  
form of the  \adss  metric is 
\be\label{cfc}
ds^2 =R^2 Z^{-2}(dx^a dx^a + dZ^MdZ^M) \ , \ \ \ \ \ \ \ \ \ 
 Z^M = Y^{-2}  Y^M  = e^{-\p} u^M  \ . 
\ee
The full superstring Lagrangian expressed in terms of 
these    coordinates  was given in \ci{MT3} and  in Appendix C in \ci{MTT}.
Its phase space counterpart  has the same structure 
 as \rf{laah},\rf{haha} 
 \ci{MTT}
\be
{\cal L} = \PP_\al\dot{x}_\al 
+ \PP_M \dot{Z}^M
+\frac{{\rm i}}{2}p^+(\theta^i \dot{\theta}_i
+\eta^i\dot{\eta}_i - h.c.) - {\cal  H}  \ , \la{lol} \ee 
$$ {\cal  H} = 
\frac{1}{2p^+}\Bigl[\PP_\al^2 +\PP_M\PP_M
+ \T^2 Z^{-4}(\x'_\al^2+ \Z'^M\Z'^M)
$$  $$
+ \  Z^{-2}[p^{+2}(\eta^2)^2 + 2 {\rm i}p^+\eta \rho^{MN} \eta  Z_M \PP_N]\Bigr]
  $$  \be \la{zez} 
- \  \T \Bigl[\ |Z|^{-3}\eta^i \rho_{ij}^M Z^M
(\th'^j - {\rm i}\sqrt{2}|Z|^{-1} \eta^j\x')+h.c.\Bigr] \ , 
\ee 
with  
$p^+\x'^-  + \PP_\al\x'_\al 
+ \PP_M \Z'^M
+\frac{{\rm i}}{2}p^+(
\theta^i\th'_i+\eta^i\et'_i - h.c.) =0$. 
Eliminating the momenta $\PP_\al$ and $ \PP_M$ 
 in \rf{lol} gives the following  \lc Lagrangian 
corresponding to  the Hamiltonian density \rf{zez} 
$$
{\cal L}  =
{  p^+ \ov 2} \bigg[ \dot{x}_\al^2  + ( \dot Z^M  - {\rm i} 
\eta_i  \rho^{MN}{}^i_j \eta^j  Z_N  Z^{-2} )^2 
+  {\rm i} (\theta^i \dot{\theta}_i
+\eta^i\dot{\eta}_i - h.c.)  $$  $$ 
-  \ Z^{-2} (\eta^2)^2 
- \  \T^2 Z^{-4} (p^+)^{-2} (\x'_\al^2+ \Z'^M\Z'^M) \bigg] 
$$ \be \la{lool}   
- \  \T  \Bigl[\ |Z|^{-3}\eta^i \rho_{ij}^M Z^M
(\th'^j - {\rm i}\sqrt{2}|Z|^{-1} \eta^j\x')+h.c.\Bigr] \ . 
\ee 
As in \rf{lamm}, the  factors of $p^+$ can be absorbed into the rescaling  
$\s\to \s/p^+$ in the  action. 
The   $\T\sim \sqrt \l \to 0$ limit of this Lagrangian is 
obtained by omitting the $\s$-derivative terms. Our final result is thus 
 $$ 
{\cal L}_{\l \to 0} 
  =
{  p^+ \ov 2} \bigg[ \dot{x}_\al^2 +  {\rm i} (\theta^i \dot{\theta}_i
+ \theta_i \dot{\theta}^i )   $$
\be \la{oops}
 + \  ( \dot Z^M  - {\rm i} 
\eta_i  \rho^{MN}{}^i_j \eta^j  Z_N  Z^{-2} )^2  + 
 {\rm i} (\eta^i \dot{\eta}_i
+ \eta_i \dot{\eta}^i ) - Z^{-2} (\eta^2)^2  \bigg]  \ .  
\ee 
This  Lagrangian 
 corresponds to the $\T\to 0$ limit of the Hamiltonian 
\rf{zez} or   \rf{hahao}.
An interesting feature of \rf{oops} 
is that the  two transverse 4-d  coordinates $x_\perp$ and 
the 4 complex fermions  $ \theta_i$  corresponding to the linear part of 
the \lc $\N=4$ supersymmetry  decouple from the rest  of the variables. 

As already mentioned above,   \rf{oops} contains 
the action of the superparticle in \adss  as its part 
depending  only on 
constant modes in  $\s$. 
The important next problem is to show  that in addition to the
type IIB \adss  supergravity 
multiplet, whose presence is implied  by the correspondence 
 with  the superparticle action \ci{met3} and is expected since 
the  masses of ``protected'' states do not depend on $\l$, 
the   string spectrum following from  quantization of 
  \rf{oops} contains also massless higher spin particles in $AdS_5$. 
Their \lc description in $AdS_d$ was  given in \ci{met2,met4}: 
in the case of totally symmetric field $\p_s$  one starts  with 
totally symmetric traceless  $SO(d-2)$ tensor $\p_{I_1....I_s} $ and
 decomposes it further 
into irreducible representations of $SO(d-3)$ as
 $ \sum^s_{s'=0} \p_{s'}, \ \p_{s'}= ( \p_{i_1 ...i_{s'}})$. 
 Then, in the parametrization  of AdS space as in 
\rf{cfc} and after an appropriate rescaling, 
the fields $\p_{s'}$ satisfy the following free equations 
$ (\del^2_x + \del_Z^2  - Z^{-2}    A_{s'} ) \p_{s'} =0$. 
Here $ A_{s'} = ( s' + { d-5 \ov 2})^2 -  { 1 \ov 4} $, 
i.e.  $ A_{s'} = s'^2 -  { 1 \ov 4} $ in the case of $AdS_5$.
These equations should appear as the mass shell conditions for the 
``lightest'' string states corresponding to  
\rf{hahao} and \rf{oops}.\foot{The  non-normalizable  and normalizable solutions 
for $\p_{s}(x,z)$ which should 
represent the conformal higher spin fields and the associated 
conformal higher spin currents of the boundary theory  have indeed 
the right integer dimensions \ci{met2}:
 $\Delta= 2-s$ and $\Delta =s + d-3=s+2$.}

To conclude,  we expect that 
the   \lc  approach developed in \ci{met2,met3} and \ci{MT3,MTT} should
be instrumental  in proving the conjectured
AdS/CFT correspondence \ci{sund,witt}
in the  $\l\to 0$ limit.

\section*{Acknowledgments}
This  work  was  supported  in part by
the DOE grant  DE-FG02-91ER-40690,    
  PPARC  grant  PPA/G/S/1998/00613,   INTAS project 
 991590 and CRDF RPI-2108  grant. 
We are grateful to   S. Frolov, I. Klebanov, A. Mikhailov, 
   H.  Nastase,   W. Siegel,   M. Vasiliev and especially  R. Metsaev 
  for useful discussions.


\end{document}